\documentclass[letterpaper, 10 pt, conference]{ieeeconf}  
\usepackage{xcolor}
\usepackage{siunitx}
\usepackage{booktabs}
\usepackage{mathtools}
\usepackage{amssymb}
\usepackage{cite}
\usepackage{graphicx}
\usepackage{epstopdf}

\IEEEoverridecommandlockouts                              

\title{\LARGE \bf
	A Novel Entropy-Maximizing TD3-based Reinforcement Learning for Automatic PID Tuning 
}

\author{Myisha A. Chowdhury$^{1}$ and Qiugang Lu$^{1,\dagger}$
	\thanks{*This work was supported by the Texas Tech University.}
	\thanks{$^{1}$M.A. Chowdhury and Q. Lu are with the Department of Chemical Engineering, Texas Tech University, Lubbock, TX 79405, USA.   
		Email: {\tt\small myisha.chowdhury@ttu.edu; jay.lu@ttu.edu}}%
	\thanks{$^{\dagger}$Corresponding author: Q. Lu }%
}

\begin{document}

	\maketitle
	\thispagestyle{empty}
	\pagestyle{empty}

\begin{abstract}
Proportional-integral-derivative (PID) controllers have been widely used in the process industry. However, the satisfactory control performance of a PID controller depends strongly on the tuning parameters. Conventional PID tuning methods require extensive knowledge of the system model, which is not always known especially in the case of complex dynamical systems. In contrast, reinforcement learning-based PID tuning has gained popularity since it can treat PID tuning as a black-box problem and deliver the optimal PID parameters without requiring explicit process models. In this paper, we present a novel entropy-maximizing twin-delayed deep deterministic policy gradient (EMTD3) method for automating the PID tuning. In the proposed method, an entropy-maximizing stochastic actor is employed at the beginning to encourage the exploration of the action space. Then a deterministic actor is deployed to focus on local exploitation and discover the optimal solution. The incorporation of the entropy-maximizing term can significantly improve the sample efficiency and assist in fast convergence to the global solution. Our proposed method is applied to the PID tuning of a second-order system to verify its effectiveness in improving the sample efficiency and discovering the optimal PID parameters compared to traditional TD3. 
\end{abstract}

\section{Introduction}
Modern process industry is characterized by high complexity due to strong coupling among different units and thus efficient control is critical for maintaining high-level operations. Among various process control strategies, PID control has received widespread attention owing to its simplicity (with only three tunable parameters) in design and effectiveness in delivering high control performance for numerous real-world applications \cite{wang2007proposal}. However, these PID parameters require careful tuning for yielding superior control performance. One class of naive PID tuning methods is based on the trial-and-error approach, where different PID parameters are tested and the one giving the best control performance is deployed to the system. However, this trial-and-error approach is time-consuming and may result in sub-optimal PID parameters \cite{dogru2022reinforcement}. Another category of PID tuning techniques is the rule-based methods where the models of the process are often required \cite{seborg2016process}. However, such models may not be always available, especially for complex processes, which restricts the applicability of these methods. To address this issue, data-driven techniques based on, e.g., evolution optimization \cite{nyberg2017optimizing}, genetic algorithm \cite{mitsukura1999design}, and neural networks (NNs) \cite{d2007use}, have been developed for PID tuning without requiring a process model. However, these methods need a large quantity of labeled data for training due to their model-free nature, making them sample inefficient \cite{guan2021design}. 

Recently, reinforcement learning (RL) has seen a surge in popularity in the control of dynamical systems \cite{zheng2020constrained}. RL is essentially a sequential decision-making process for optimizing a black-box objective function  \cite{sutton2018reinforcement}. Unlike supervised learning, model-free RL can learn the best policy to optimize the objective function from direct interactions with the environment and does not require prior knowledge about the environment. On the other hand, the PID tuning can be seen as a black-box optimization problem where the relation between PID parameters and the resultant control performance is unknown. On this aspect, RL has great potential to be an effective technique for solving the PID tuning problem through the sequential decision-making process \cite{lakhani2021stability}. 

In light of these observations, research studies have been reported on using RL-based algorithms for PID tuning \cite{wang2007proposal,dogru2022reinforcement,guan2021design,zheng2020constrained,lakhani2021stability, younesi2019q, shi2020adaptive, lawrence2020reinforcement}. For example, Q-learning, a popular RL algorithm, has been used to design PID controllers that are adaptive to system operating condition changes \cite{younesi2019q}. However, the Q-learning algorithm works only on discrete state-action spaces, and cannot handle the continuous space \cite{shi2020adaptive}. One solution to this problem is to discretize the states and actions into bins before applying Q-learning \cite{lawrence2020reinforcement}. Nonetheless, such discretizations may lead to errors due to the finite resolution, and eventually result in poor control performance \cite{shi2020adaptive, noel2014control}. To this end, actor-critic-based algorithms have been utilized where the policy and Q-values are approximated by parameterized functions \cite{sutton2018reinforcement,wang2007proposal,dogru2022reinforcement}. In particular, the deep deterministic policy gradient (DDPG) algorithm proposed by DeepMind \cite{silver2014deterministic}, which is sample-efficient and effective in handling continuous state-action space, has been adapted to the PID tuning problem \cite{lakhani2021stability}. An updated version of DDPG with lower variance, known as twin delayed DDPG (TD3), has also been applied to PID tuning for nonlinear systems \cite{shi2020adaptive}. Although the TD3 algorithm has demonstrated improved sample efficiency compared to the other existing methods such as DDPG \cite{fujimoto2018addressing}, it often performs poorly in high-dimensional state-action space due to its inherent lack of exploration \cite{haarnoja2018soft}, which limits its efficacy in PID tuning. 

In this article, we present a novel algorithm for facilitating the PID tuning, termed as entropy-maximizing TD3 (EMTD3), to overcome the issue of insufficient exploration associated with the traditional TD3 algorithm. Specifically, the developed EMTD3 algorithm deploys an entropy-based stochastic actor to ensure significant exploration at the beginning, followed by a deterministic actor based on the TD3 algorithm. For our method, the stochastic actor can well ensure sufficient explorations initially and the deterministic actor can facilitate fast convergence once knowledge about the environment is acquired. Moreover, the off-policy nature of the EMTD3 algorithm enables it to use past experience and makes this method sample efficient. The proposed method is applied to the PID tuning of a second-order system. Simulation results show that our method outperforms traditional TD3 significantly in improving the sampling efficiency and discovering optimal PID parameters. 

This paper is organized as follows. Section \ref{sec: preliminaries} presents fundamentals about PID control and the TD3 algorithm, followed by our proposed EMTD3 method elaborated in \ref{sec: EMTD3}. The numerical case study is presented in Section \ref{sec: simulation} with our conclusions given in Section \ref{sec: conclusion}. 

\section{Preliminaries}
\label{sec: preliminaries} 
\subsection{PID Control}

The typical form of a PID controller has three adjustable parameters and can be expressed as \cite{seborg2016process}:
\begin{equation}
		u(t_{n}) = K_{p}\left[e(t_{n}) + \frac{1}{\tau_{I}} \sum_{i=1}^{n}e(t_i) + \tau_{D} \frac{e(t_n)-e(t_{n-1})}{\Delta t}\right], \label{eq: PID}
\end{equation}
where $t_n$ is the $n$-th time instant, $u(t_n)$ is the manipulated variable (MV), $e(t_n)$ is the error between the setpoint $y_{sp}$ and the controlled variable (CV) $y(t_n)$, and $\Delta t$ is the sampling interval. $K_p$, $\tau_I$, and $\tau_D$ are the PID parameters. In this work, we use the following anti-reset windup to accommodate the potential windup issue from the integral part  \cite{lakhani2021stability}:  
\begin{equation}
	sat(u)=
	\begin{dcases}
		u_{min}, \quad & \text{if} \quad  u < u_{min}, \\
		u, \quad & \text{if} \quad  u_{min} \le u \le u_{max}, \\
		u_{max}, \quad & \text{if} \quad  u > u_{max}, 
	\end{dcases} \label{eq: Saturation}
\end{equation}
where $u_{min}$ and $u_{max}$ are the lower and upper bounds of the MV, respectively.

\subsection{Reinforcement Learning}
RL formulates the optimal control problem as a Markov decision process expressed as $(\mathbb{S},\mathbb{A},\mathbb{P},R)$, where $\mathbb{S}$ is the state space, $\mathbb{A}$ is the action space, $\mathbb{P}$ is the transition probability matrix, and $R$ is the reward signal \cite{sutton2018reinforcement}. In RL, an agent observes the state $s_t$ and deploys an action $a_t$ under current policy to the environment. The environment then evolves one step forward under $a_t$, generating a reward signal $r_t:=R(s_t, a_t)$  to the agent. To assess the action $a_t$ taken at state $s_t$, the Q-value $Q^{\pi}(s_t, a_t)$ is defined as the expected sum of future discounted rewards from time $t$ onward:
\begin{equation}
	Q^{\pi}(s_t,a_t) = \mathbb{E}\left[\sum_{t=0}^{\infty}\gamma^tr_{t+1} | s=s_t, a=a_t,\pi \right], \label{eq: q_value}
\end{equation}    
where $\gamma\in[0,1]$ is the discount factor and $\pi(a_t|s_t):s_t\to a_t$ is the underlying policy mapping state $s_t$ to action $a_t$. The objective of the RL problem is to maximize the total cumulative reward by finding the optimal policy $\pi^{*}(a_t|s_t)$ from the optimal Q-value $Q^*(s_t, a_t)$
\begin{equation}
	\pi^{*}(a_t|s_t)=\begin{dcases}
		1, \quad &\text{if} \quad a_t=\arg \max_{a_t\in\mathbb{A}}~ Q^{*}(s_t, a_t),\\
		0, \quad &\text{otherwise}.
	\end{dcases}   \label{eq: optimal_policy}
\end{equation}
For a deterministic policy $\mu(a_t|s_t)$, the Q-value can be described by the Bellman expectation condition as
\begin{equation}
Q^{\mu}(s_t, a_t)=\mathbb{E} \left[ r_t +  \gamma Q^{\mu}(s_{t+1}, \mu(s_{t+1}))\right]. \label{eq:DeterministicQ}
\end{equation}
For a high-dimensional RL problem, it is challenging to learn the Q-value for each state-action pair individually. Under such circumstances, a function approximator (FA), generally a NN $Q_\theta(s_t,a_t)$, parameterized by $\theta$, is used to learn the true Q-value. The objective of the FA is to minimize the approximation error by finding the optimal parameter $\theta^{*}$: 
\begin{equation}
\theta^{*}=\arg\min_{\theta}~\mathbb{E}\left[(y_t-Q_\theta(s_t,a_t))^2\right], \label{eq:CriticLoss}
\end{equation}
where $y_t= r_t+\gamma \max_{a}Q_\theta(s_{t+1},a)$ is the temporal-difference (TD) target.

In addition to the value-based method above, another important RL category is policy-based methods, where the RL  searches directly the optimal policy $\mu_\phi(s_t)$, parameterized by $\phi$, to maximize the accumulated reward $J(\mu_\phi)=\mathbb{E}_{\mu_\phi}\left[\sum_t R(s_t,a_t)\right]$. This class of methods uses gradient ascent to update $\phi$  \cite{lakhani2021stability} 
\begin{equation}
\phi_{t+1}\leftarrow \phi_{t} + \alpha \nabla_{\phi} J(\mu_\phi)|_{\phi=\phi_t}, \label{eq: PG_update}
\end{equation}
where $\alpha$ is the learning rate, and the gradient of the cost $J(\mu_\phi)$, using the policy-gradient theorem \cite{sutton2018reinforcement}, is 
\begin{equation}
\nabla_{\phi} J(\mu_\phi) = \mathbb{E}_{\mu_\phi}\left[\nabla_{a} Q^{\mu}(s_t,a)|_{a=\mu_{\phi}(s_t)}\nabla_{\phi}\mu_{\phi}(s_t)\right]. \label{eq: DPG_gradient}
\end{equation}

The actor-critic method combines both value-based and policy-gradient methods where the actor generates actions depending on the current state and the critic evaluates the action that was taken \cite{wang2007proposal, dogru2022reinforcement, lakhani2021stability}.
  
\subsection{DDPG and TD3}
As a model-free off-policy actor-critic method, the DDPG framework employs a deterministic policy and thus is sample-efficient \cite{silver2014deterministic}. The application of DDPG to PID tuning has been reported in \cite{lakhani2021stability}. However, the critic network in the DDPG suffers from the issue of overestimation bias due to the maximizing operation for calculating the TD target in \eqref{eq:CriticLoss}. To address this problem, the TD3 method is proposed where two different target Q networks, parameterized by $\theta_i^{\prime}$, $i=1,2$, are used for estimating two Q-values and the smaller one is then used for computing the TD target \cite{fujimoto2018addressing}:
\begin{equation}
y_t= r_t+\gamma \min_{i=1,2} Q_{\theta_i ^\prime}(s_{t+1},\mu_{\phi^{\prime}}(s_{t+1})), \label{eq:q_target}
\end{equation} 
where $\phi^{\prime}$ is the parameter of the target actor network. Moreover, the parameters of the actor and target networks are updated less frequently to reduce the accumulation of function approximation error and hence reduce the variance of Q-value estimate. 
In addition, a clipped noise is added to the target action to enable better exploration by adding uncertainty to the policy, further reducing the variance. Then the new target value is formulated as \cite{fujimoto2018addressing}
\begin{align}
y_t= & r_t+\gamma   Q_{\theta^\prime}(s_{t+1},\mu_{\phi^\prime}(s_{t+1})+\epsilon),
\end{align}
where $\epsilon\sim clip(\mathcal N (0,\sigma), -c, +c)$ is the added Gaussian noise, $\theta^\prime$ and $\phi^\prime$ are the parameters of the selected target critic and  actor networks giving smaller Q-value in \eqref{eq:q_target}, and $c$ is the bound on the target action. Details on the DDPG and TD3 algorithms can respectively be found in \cite{silver2014deterministic}, \cite{fujimoto2018addressing}.
                 
\section{The Proposed EMTD3 for PID Tuning}
\label{sec: EMTD3}
\subsection{Entropy-Maximizing TD3 (EMTD3)}
In this paper, we propose a novel hybrid RL algorithm, termed as EMTD3, consisting of a stochastic actor for exploration and a deterministic actor for exploitation. At the initial stage, an entropy regularization term is added to the RL objective to simultaneously maximize the entropy along with the total cumulative reward:
\begin{equation}
J(\pi)=\sum_{t=0}^{T}\mathbb{E}_{(s_t,a_t)\sim\rho_{\pi}}[r_t+\beta\mathbb{H}(\pi(\cdot|s_t))], \label{eq:em_rl}
\end{equation} 
where $\pi$ is the stochastic policy, the entropy  $\mathbb{H}(\pi(\cdot|s_t))=-\int_{a_t}\pi(a_t|s_t)\log\pi(a_t|s_t)$,  $\rho_{\pi}$ is the state-action distribution, and $\beta$ is the parameter controlling the relative weight between the two terms. The added entropy-maximizing term encourages the agent to sufficiently explore the state-action space and avoid convergence to the local solution. With the RL objective in \eqref{eq:em_rl}, based on the Bellman equation, the new Q-value can be obtained by expanding \eqref{eq: q_value} into 
\begin{align}
Q^\pi(s_t,a_t)=& \sum_{t=0}^{T}\mathbb{E}_{(s_t,a_t)\sim\rho_{\pi_{\phi_1}}}[r_t+\nonumber\\&\gamma(Q^\pi(s_{t+1},a^{\prime})-\beta\log \pi(a^{\prime}|s_{t+1}))], \label{eq:q_em_rl_2}
\end{align}
where $a^{\prime}\sim \pi(\cdot|s_{t+1})$ and is sampled from $\pi$. As discussed in Section \ref{sec: preliminaries}, in the continuous space, we use FAs, parameterized by $\theta$ and $\phi_1$, respectively, to estimate the  Q-value and policy. The optimal parameter $\theta^{*}$ in the FA for the Q-value can be obtained by minimizing the loss function 
\begin{equation}
\theta^{*}=\arg\min_{\theta}~\mathbb{E}\left[(y_t-Q_\theta(s_t,a_t))^2\right].
\end{equation}
Similar to the deep Q network, we compute the TD target $y_t$ by using a separate NN, parameterized by $\theta^{\prime}$,
\begin{align}
y_t= r_t+ \gamma (Q_{\theta ^\prime}(s_{t+1},a^{\prime})-\beta \log \pi_{\phi_{1}}(a^{\prime }|s_{t+1})).
	\label{eq:target_em}
\end{align}
The parameter $\phi_1$ of the stochastic policy can be calculated utilizing the re-parameterization trick introduced in \cite{haarnoja2018soft}.  
\begin{figure} [thpb]
	\centering
	\includegraphics[width=\columnwidth]{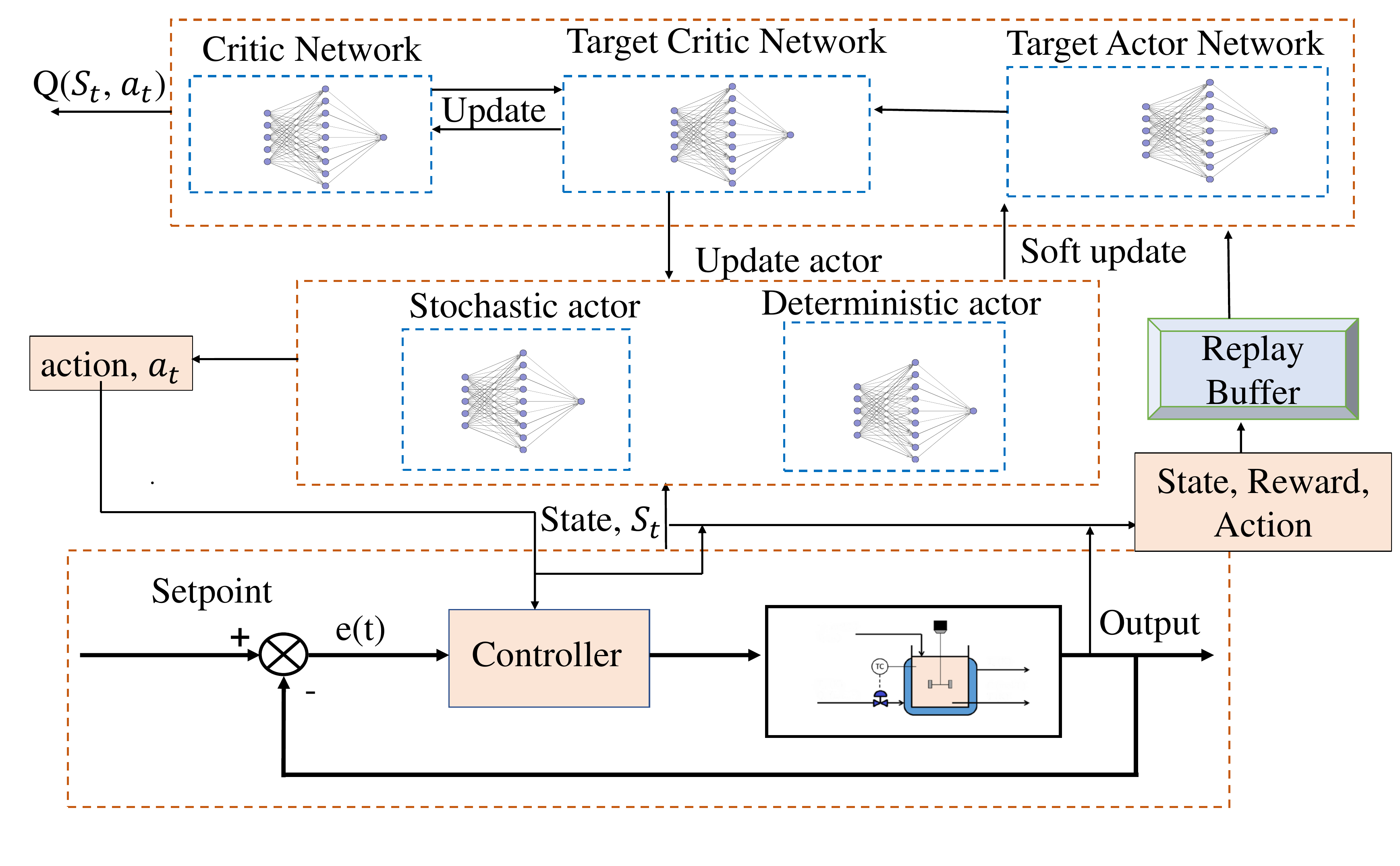}
	\caption{The structure of automatic  PID tuning on the  EMTD3 framework.}
	\label{fig: EMTD3}
\end{figure}
Adding the entropy-maximizing term to the RL objective assists the stochastic policy in exploring the action space. However, it may result in  overestimation of the entropy and hence demonstrate unstable behaviors during training \cite{ding2021averaged,ali2021reducing}. In addition,  stochastic policy methods often require a large number of samples for training in continuous action space and may show slow convergence. On the other hand, deterministic policy methods do not suffer from above issues. However, the critic networks in deterministic methods often generate peaks for some actions and remain stuck in that region without finding the global optimum \cite{ali2021reducing}.

In the proposed EMTD3 method, we combine entropy-maximizing stochastic policy and deterministic policy together to preserve the respective advantages. Fig. \ref{fig: EMTD3} shows the structure of the proposed EMTD3 algorithm. It consists of two critic networks, two target critic networks, a deterministic, a stochastic, and a target actor network. The developed EMTD3 method is divided into two stages. For the first stage (warm-up period), we implement the entropy-maximizing stochastic actor focusing on the exploration, which reduces the action peaks by adding uncertainty and assists in effectively exploring the action space \cite{ali2021reducing}. After the first stage, the policy shifts to a deterministic one (TD3) to focus on exploitation and accelerate the convergence. Note that both stages share the same critic networks and critic target networks, but the actor networks differ. In other words, once the first stage is complete, these critic networks and their target networks will have been preliminarily trained. This can greatly reduce the variance associated with Q-value estimates in the second stage, which is beneficial for speeding up the learning process of the deterministic actor. In addition, we use (\ref{eq:target_em}) and (\ref{eq:q_target}) respectively to calculate the TD target $y_t$ for stochastic and deterministic policies, where two different target networks are used to compute two Q-values and the smaller one is selected to alleviate the overestimation error. The parameters of the critic networks can be updated by using the gradient of the following cost function that quantifies the approximation error:
 \begin{equation}
 J(Q_{\theta_i})=\frac{1}{|B|} \sum\left(y_t-Q_{\theta_{i}}(s,a) \right)^2,
 \end{equation} 
where $i=1,2$ represents $i$-th critic network and $B$ is the batch size sampled from the replay buffer that stores the experience $(s_t,a_t,r_t,s_{t+1})$. The parameters of the stochastic and deterministic actor networks ($\phi_1$ and $\phi_2$, respectively) can be updated using the gradient of the following costs: 
\begin{align}
	J(\pi_{\phi_{1}})=& \frac{1}{|B|} \sum_{s \in B}(\min_{i=1,2}Q_{\theta_i}(s,a_{\phi_{1}}^{\prime }(s))-\nonumber\\&\beta \log \pi_{\phi_{1}}(a_{\phi_1}^{\prime}(s)|s)), \label{eq:em_update}
\end{align} 
\begin{equation}
	J(\mu_{\phi_{2}})=\frac{1}{|B|} \sum_{s \in B}Q_{\theta_1}(s,\mu_{\phi_2}(s_t)). \label{eq:td3_update}
\end{equation}
To reduce the variance, we update the deterministic actor and target networks less frequently than the critics \cite{fujimoto2018addressing}. The target network parameters are updated by ($\rho$ is close to 1):
\begin{align}
\phi_2^\prime\leftarrow \rho\phi_2^\prime+(1-\rho)\phi_2,~ \theta_{i}^\prime\leftarrow \rho \theta_{i}^\prime+(1-\rho)\theta_{i}.
\end{align}

\begin{table*}[tbh]
	\centering
	\caption{Pseudo code for EMTD3-based PID tuning} \label{Algo_EMTD3_PID_Tuning}
	\begin{tabular}{ll}
		\hline
		\multicolumn{2}{l}{\textbf{Algorithm: EMTD3 for PID tuning}} \\ \hline
		1: & \textbf{Input:} initial Q-value FA parameters $\theta_1$, $\theta_2$, stochastic and deterministic policy parameter $\phi_1$, $\phi_2$, and empty replay buffer $\mathcal{D}$ \\
		2: & Initialize target networks $\theta_1^{\prime}\leftarrow\theta_1$, $\theta_2^{\prime}\leftarrow\theta_2$, $\phi_1^{\prime}\leftarrow\phi_1$,
		$\phi_2^{\prime}\leftarrow\phi_2$ \\
		3:&\textbf{Repeat} \\
		4:& If time-step $<$ warm-up period:\\
		5:&\qquad \qquad Observe state $s$ and select action $a\sim \pi_{\phi_1}(\cdot|s)$ \\
		
		6:& If time-step $\ge$ warm-up period:\\
		7:&\qquad \qquad Observe state $s$ and select $a=clip(\mu_{\phi_2}(s)+\epsilon, a_{low}, a_{high})$ , where $\epsilon$ is random noise.\\
		8:& Execute $a$ in the environment \\
		9:& Observe the next state $s^{\prime}$ and compute the episodic reward $r$ using the closed-loop data \\
		10:& Store the transition $\left(s,a,r,s^{\prime}\right)$ into the replay buffer $\mathcal{D}$ \\
		11:& If $s^\prime$ is terminal, reset environment\\
		12:&  \textbf{Update RL agent parameters:} \\
		13:&  Randomly select a batch of transitions $B=\left\{\left(s,a,r,s^{\prime}\right)\right\}$ from $\mathcal{D}$ with 
		the number of transitions as $|B|$ \\
		14: & Compute the target for the Q-value:\\
		15:& $\qquad $ If time-step $<$ warm-up period:\\
		16: & \qquad \qquad $y= r+ \gamma (\min_{i=1,2} Q_{\theta_i ^\prime}(s^\prime,a^{\prime})-\alpha \log \pi_{\phi_{1}}(a^{\prime}|s^\prime))$, where $a^{\prime}\sim \pi(.|s^\prime)$ and is sampled from $\pi_{\phi_1}$  \\
		17: & $\qquad $ If time-step $\ge$ warm-up period:\\
		18: & \qquad \qquad $y=r+\gamma (\min_{i=1,2} Q_{\theta_i ^\prime}(s^\prime,a^{\prime}(s^\prime)))$, where $a^{\prime}(s^\prime)=clip(\mu_{\phi^\prime_2}(s^\prime))+clip(\epsilon,a_{low},a_{high}))$\\
		19:& Update the Q-value by one step of gradient descent using cost function w.r.t. $\theta_i$: \\
		& \qquad \qquad  $\nabla_{\theta_i}\frac{1}{|B|} \sum_{\left(s,a,r,s^{\prime}\right)\in B}\left(Q_{\theta_{i}}(s,a) - y \right)^2$ \\
		20: &  If time-step $<$ warm-up period:\\
		21: &  \qquad \qquad Update the policy by one step of gradient descent using cost function w.r.t. $\phi_1$:\\
		22:&  \qquad \qquad $\nabla_{\phi_1}\frac{1}{|B|}\sum_{\left(s,a,r,s^{\prime}\right) \in B}(\min_{i=1,2}Q_{\theta_i}(s,a_{\phi_{1}}^{\prime }(s))-\alpha \log \pi_{\phi_{1}}(a_{\phi_1}^{\prime }(s)|s))$, where $a_{\phi_1}^{\prime}(s)$   is sampled from $\pi_{\phi_1}(\cdot|s)$\\
		23: &\qquad \qquad Update the parameters $\theta_i^{\prime}$  in the target networks\\
		
		24: &  \qquad \qquad $\theta_{i}^\prime\leftarrow \rho \theta_{i}^\prime+(1-\rho)\theta_{i}$\\
		25: & If  time-step $\ge$ warm-up period: \\
		
		26: &  \qquad  If $time-step \% d == 0$ where $d$ is the update period, then\\
		27: &  \qquad \qquad$\nabla_{\phi_2}\frac{1}{|B|}\sum_{\left(s,a,r,s^{\prime}\right) \in B}Q_{\theta_1}(s, \mu_{\phi_2}(s))$\\
		28: &  \qquad \qquad  Update target networks with:\\
		29: &  \qquad \qquad  $\theta_{i}^\prime\leftarrow \rho \theta_{i}^\prime+(1-\rho)\theta_{i}$\\
		30:&  \qquad \qquad  $\phi_2^\prime\leftarrow \rho\phi_2^\prime+(1-\rho)\phi_2$\\
		31:& \textbf{Until Convergence then End Repeat}  \\
		\hline
	\end{tabular}%
\end{table*}

\subsection{EMTD3 on PID Tuning}
In the proposed EMTD3-based PID  tuning framework (see Fig. \ref{fig: EMTD3}), the environment comprises the closed-loop system that can generate MVs, CVs, and setpoint from a closed-loop simulation.
For this framework, each closed-loop simulation represents \textit{one agent-environment interaction} in the EMTD3 algorithm. To reduce the dimension, we define the environment state $\mathbf{s}_k$ after the $k$-th closed-loop simulation to be a subset of the  trajectories of process variables: 
\begin{align}
&\mathbf{s}_k=\left[\tilde{\mathbf{y}}_k, \tilde{\mathbf{u}}_k, \tilde{\mathbf{y}}^{sp}_k\right]^\mathsf{T},~~\tilde{\mathbf{y}}_k=[y_{0}^{k},y_{i}^{k},y_{2i}^{k},\ldots]^\mathsf{T}, ~~\nonumber\\ &\tilde{\mathbf{u}}_k=[u_{0}^{k},u_{i}^{k},u_{2i}^{k},\ldots]^\mathsf{T}, ~~\tilde{\mathbf{y}}_k^{sp}=[y_{0}^{sp,k}, y_{i}^{sp,k}, y_{2i}^{sp,k},\ldots]^\mathsf{T},\nonumber \label{eq: Env_state}
\end{align}
where $k$ stands for the $k$-th closed-loop operation and $i$ is the sampling interval. The setpoint $\tilde{\mathbf{y}}^{sp}_k$ is included in the state to provide the agent information about the control target. In this work, we have kept the setpoint fixed throughout the closed-loop simulation. We define the PID parameters as the RL action
\begin{equation}
\mathbf{a}_k = \left[K_p, \tau_I, \tau_D\right].
\end{equation}
The reward function $r_k$ is defined as the total accumulated tracking error  throughout one closed-loop simulation: 
\begin{equation}
r_k = -\sum_{s=0}^{K} e_s^{2}, \label{eq: ISE}
\end{equation}
where $e_s = \tilde y_{s}^{sp}-\tilde y_{s}$ is the tracking error at time $s$ and $K$ is the length of the closed-loop simulation. The pseudo code of the proposed EMTD3 method for PID tuning is illustrated in Table \ref{Algo_EMTD3_PID_Tuning}.
   
\section{Simulation}
\label{sec: simulation}
In this section, we test the proposed EMTD3-based PID tuning via the following second-order process
\begin{equation}
	G(s) = \frac{0.3}{25s^2+10s+1}e^{-10s}. \label{eq: model}
\end{equation}
The MV in the simulation is bounded between $u_{min}=-20$ to $n_{max}=100$. The setpoint $y^{sp}$ is fixed at 7.5. Table \ref{table: Hyperparameters} lists the hyperparameters adopted by the EMTD3 algorithm.   
\begin{table}[h]
	\centering
	\caption{List of hyperparameters used by the EMTD3 algorithm}
	\label{table: Hyperparameters}
	\begin{tabular}{lll}
		\hline
		Parameters & Values (Case I) & Values (Case II)\\ \hline
		Learning rate $\alpha$ for actors & 0.02 & 0.02\\
		Learning rate $\beta$ for critics & 0.0005 & 0.008 \\
		Warm-up period & 70 & 100\\
		Batch size $B$ & 40 & 40 \\
		Episode length $T$ &  200 & 200\\
		Environment state dimension $n_s$ & 30 & 30 \\
		Discount factor $\gamma$ in the return & 0.99 & 0.99\\
		Polyak averaging coefficient $\rho$ & 0.006 & 0.006\\
		Initial temperature $\beta$  & 2 & 2\\
		$\frac{1}{\beta}$ incremental factor & 0.005 & 0.0001\\ 
		Initial noise variance $\sigma^2$  & 0.05 & 0.08\\
		Noise decay factor & 0.005 & 0.0045\\ \hline 
	\end{tabular}%
\end{table}

\subsection{Case Study I}
\begin{figure} [thpb] 
	\centering
	\includegraphics[width=0.8\columnwidth]{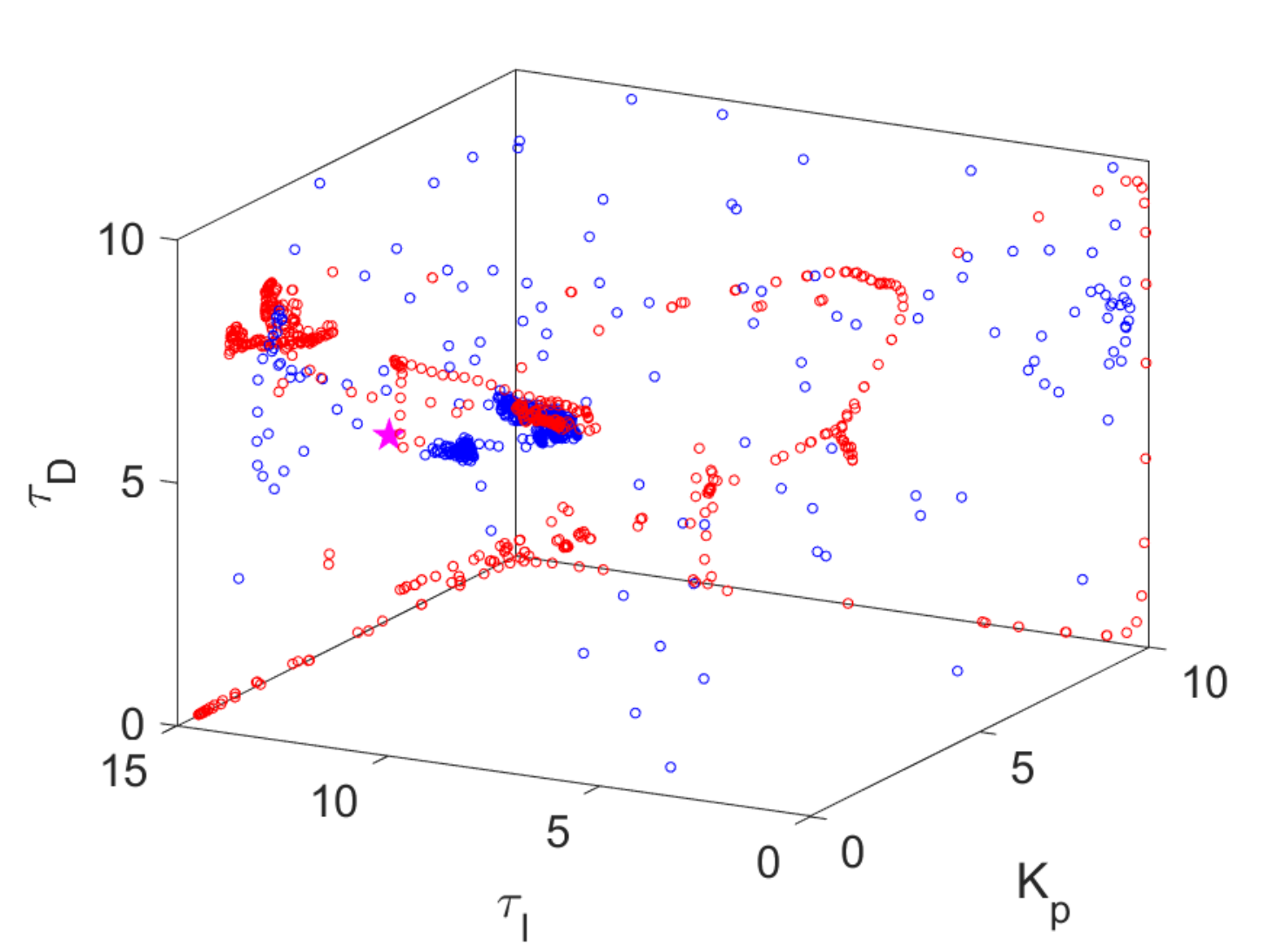}
	\caption{Exploration of the PID parameter space by TD3 (red dot) and the proposed EMTD3 algorithm (blue dot).}
	\label{fig: small_action_exploration}
\end{figure}

\begin{figure} [thpb]
	\centering
	\includegraphics[width=0.8\columnwidth]{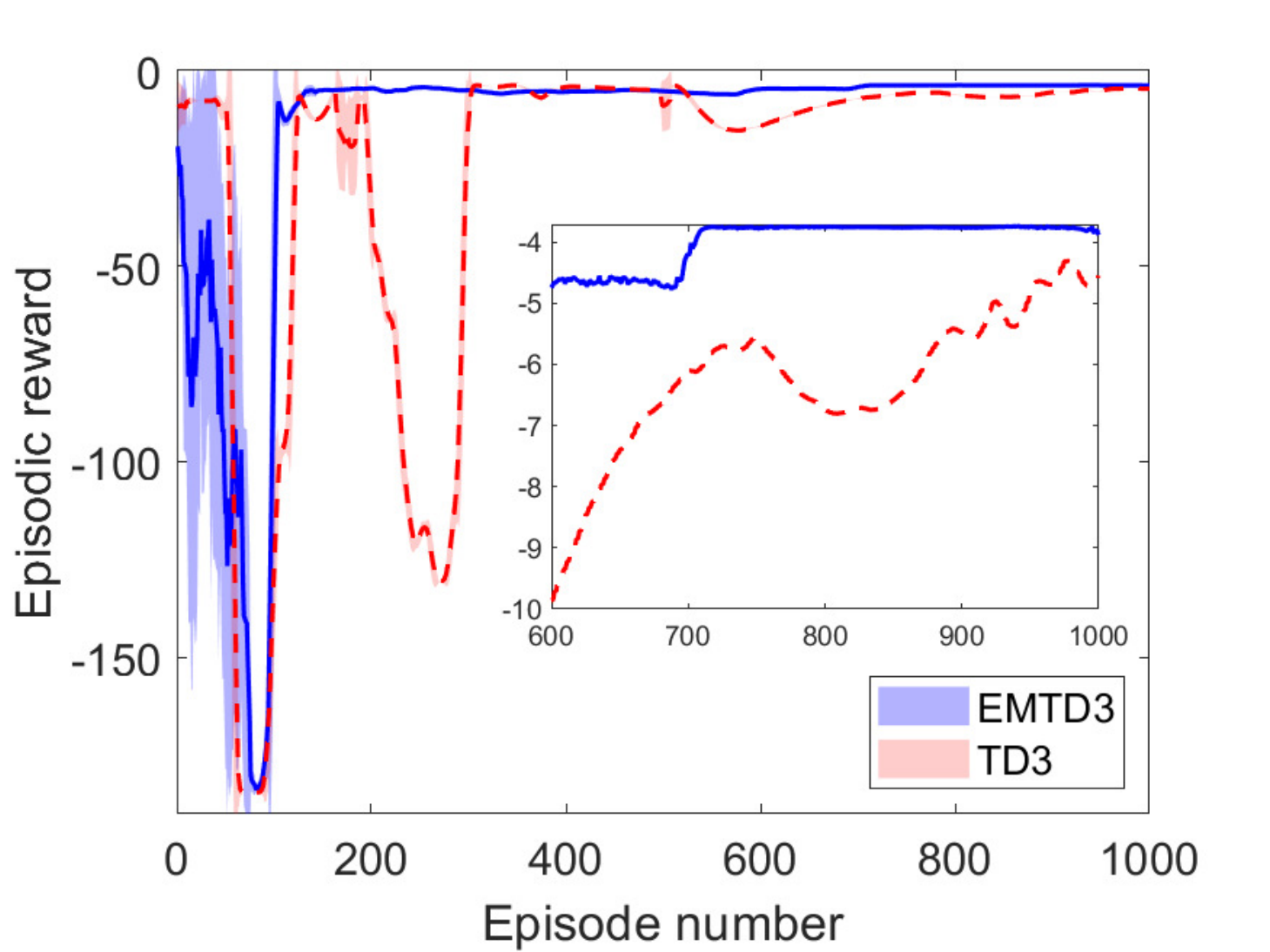}
	\caption{Learning curves with TD3 and EMTD3 for Case I.}
	\label{fig: small_action}
\end{figure}

\begin{figure} [thpb]
	\centering
	\includegraphics[width=0.8\columnwidth]{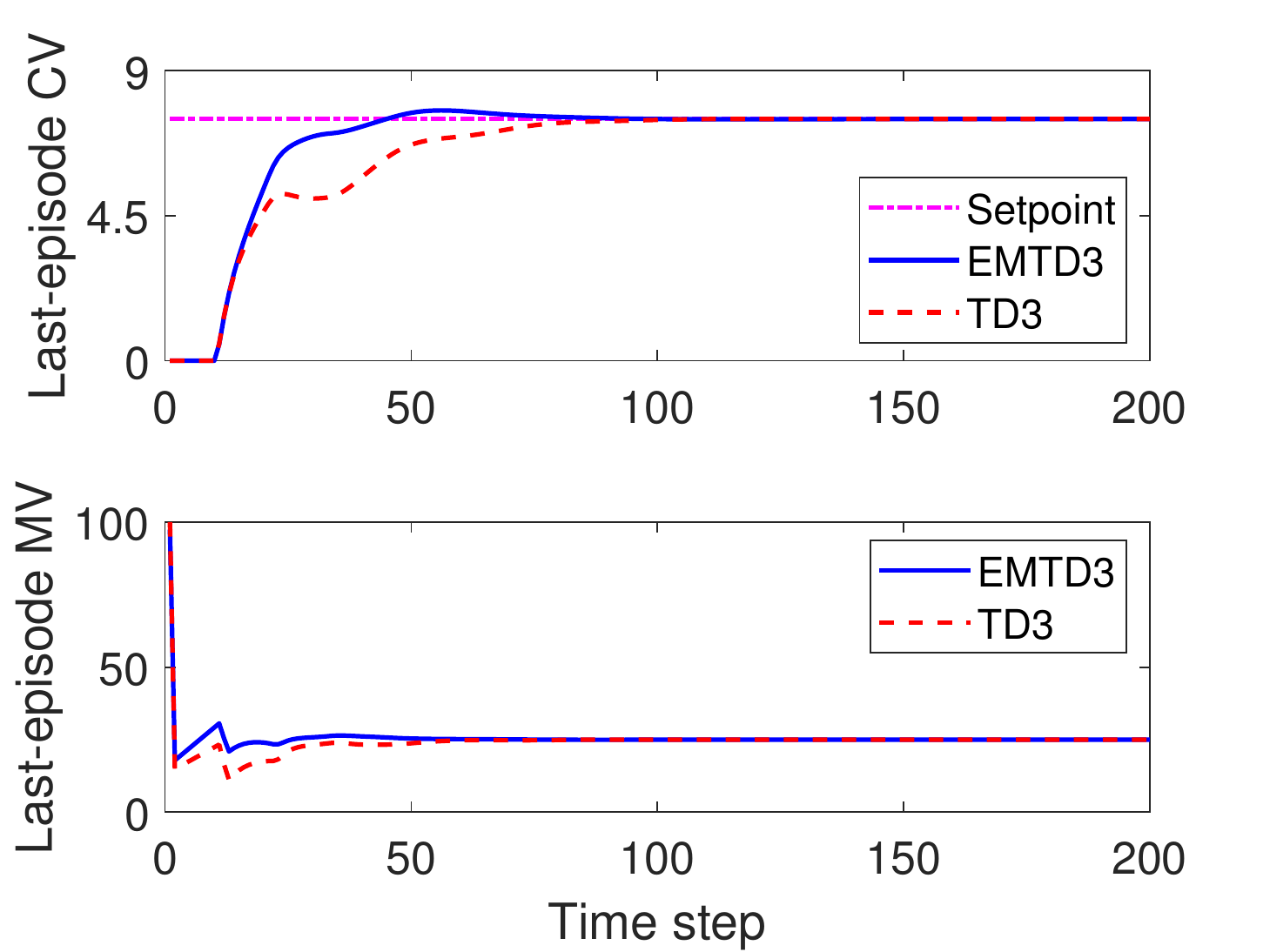}
	\caption{The PID control performance from TD3 and EMTD3 for Case I.}
	\label{fig: small_action_cv}
\end{figure}

For this case study, we specify the ranges of the PID tuning parameters to be $K_P \in [0,15]$, $\tau_{I}\in[0,15]$, and $\tau_D\in[0,10]$, respectively. The developed EMTD3 algorithm in Table \ref{Algo_EMTD3_PID_Tuning} is applied to identify the optimal PID parameters for \eqref{eq: model}. Specifically, in the warm-up stage, we gradually decrease the weight $\beta$ in \eqref{eq:em_rl} to diminish the entropy effect and thus reduce the exploration, followed by introducing a decaying zero-mean Gaussian noise after the warm-up stage. For comparison, we also use traditional TD3 with the same hyperparameters as in Table \ref{table: Hyperparameters} for the PID tuning, where a decaying zero-mean Gaussian noise is added to the action each time. Fig. \ref{fig: small_action_exploration} shows the distribution of the attempted PID parameters by the traditional TD3 (red) and our EMTD3 (blue) method, where the pentagram shows the optimal PID parameters for \eqref{eq: model}. It is observed that the TD3 method does not well explore the space and thus converges to local solutions far from the optimum. In contrast, the EMTD3 method explores the action space uniformly and converges closer to the optimum. Fig. \ref{fig: small_action} shows the learning curves between these two methods, where the EMTD3 method can settle down rapidly whereas the TD3 method needs much more training episodes. Thus, the EMTD3 show superior sample efficiency that TD3. Moreover, the discovered PID parameters from EMTD3 yield a larger reward and thus better control performance than those from TD3, which is also indicated by the step response in Fig. \ref{fig: small_action_cv}. The obtained PID parameters from EMTD3 can make the CV track the setpoint much faster without significant overshoots.  


\subsection{Case Study II}
\begin{figure} [thpb]
	\centering
	\includegraphics[width=0.8\columnwidth]{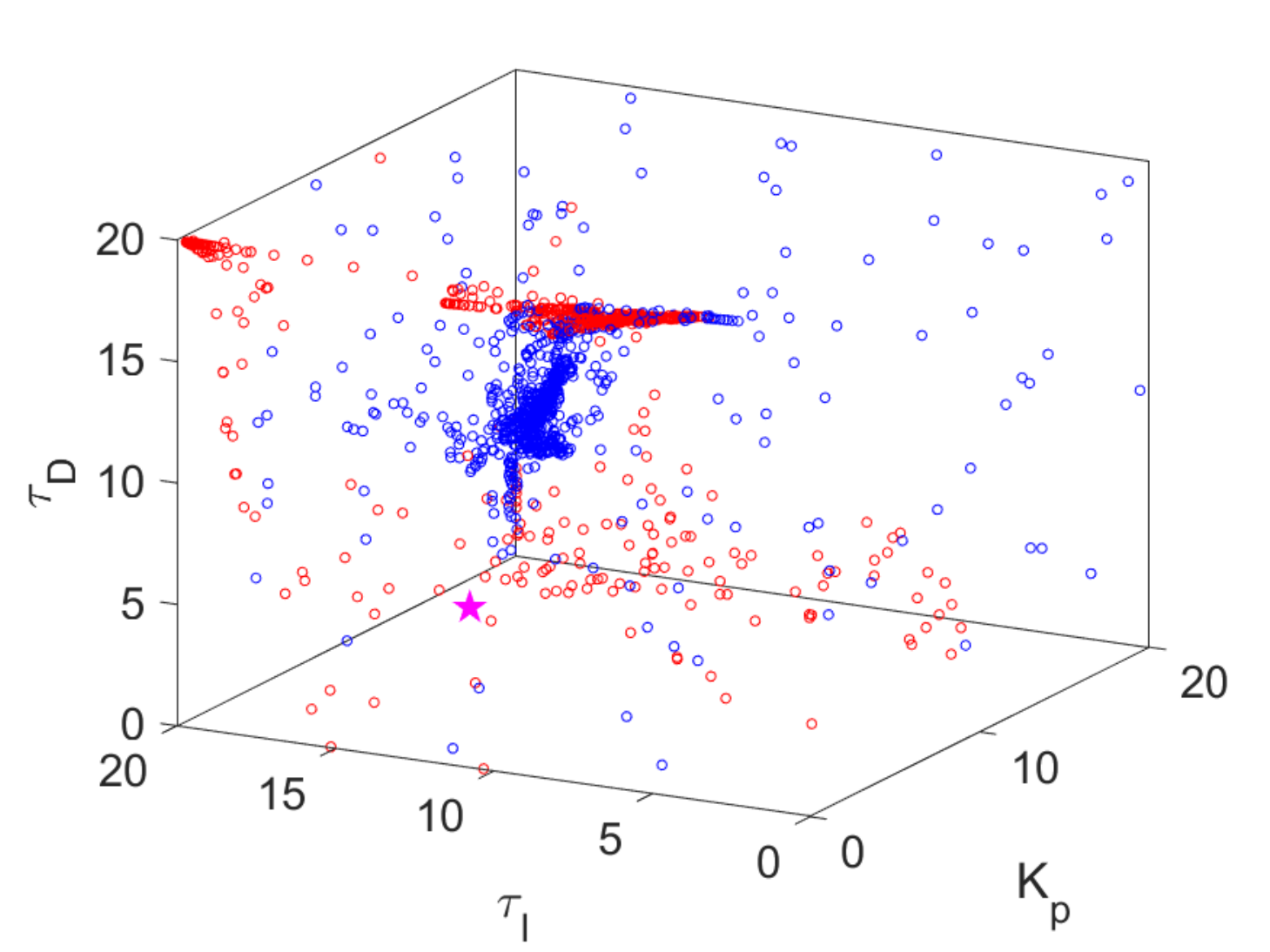}
	\caption{Exploration of the PID parameter space by TD3 (red dot) and EMTD3 (blue dot) for Case II.}
	\label{fig: medium_action exploration}
\end{figure}
\begin{figure} [thpb]
	\centering
	\includegraphics[width=0.8\columnwidth]{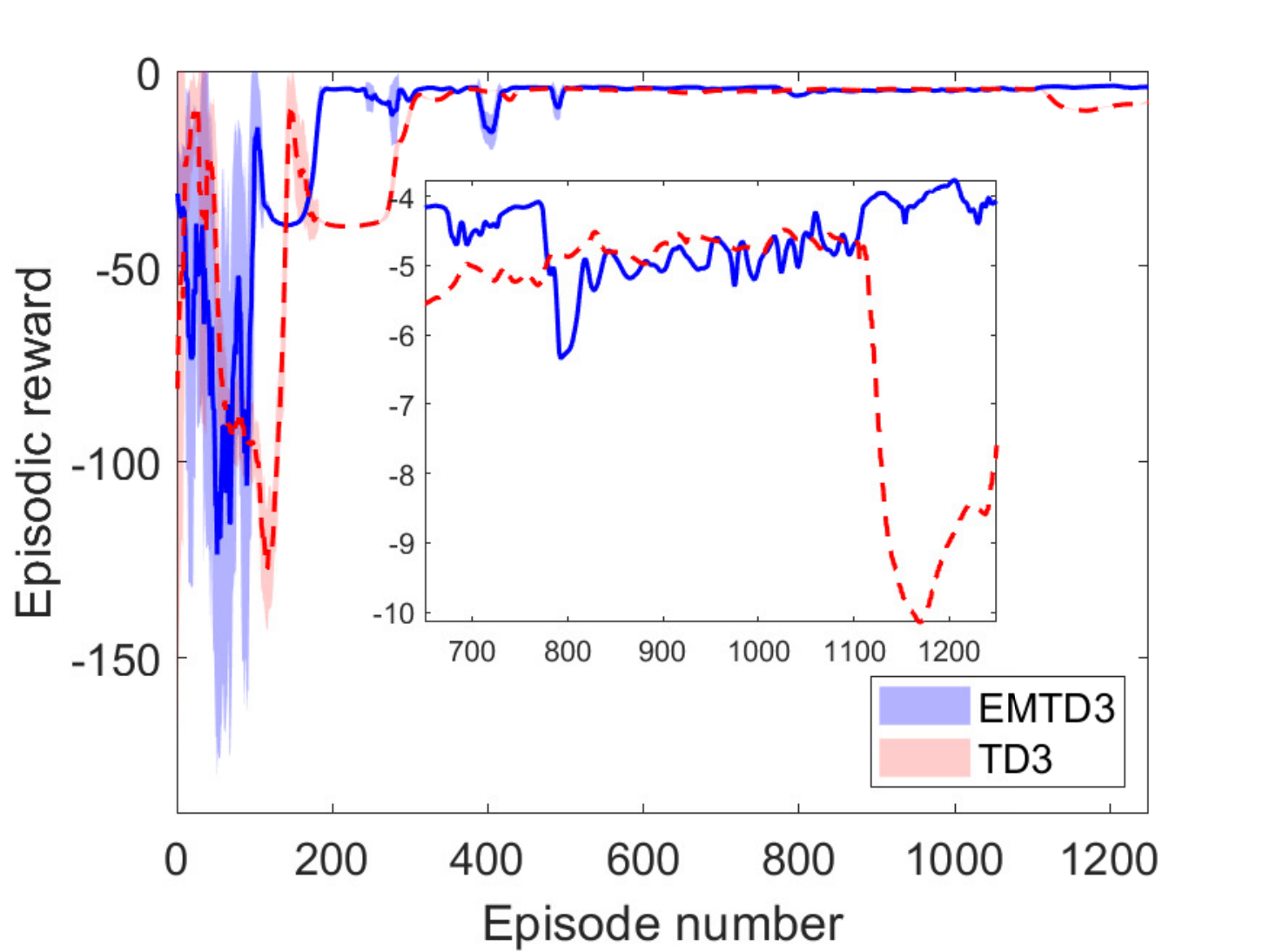}
	\caption{Learning curve with TD3 and EMTD3 for Case II.}
	\label{fig: medium_action}
\end{figure}
\begin{figure} [thpb]
	\centering
	\includegraphics[width=0.8\columnwidth]{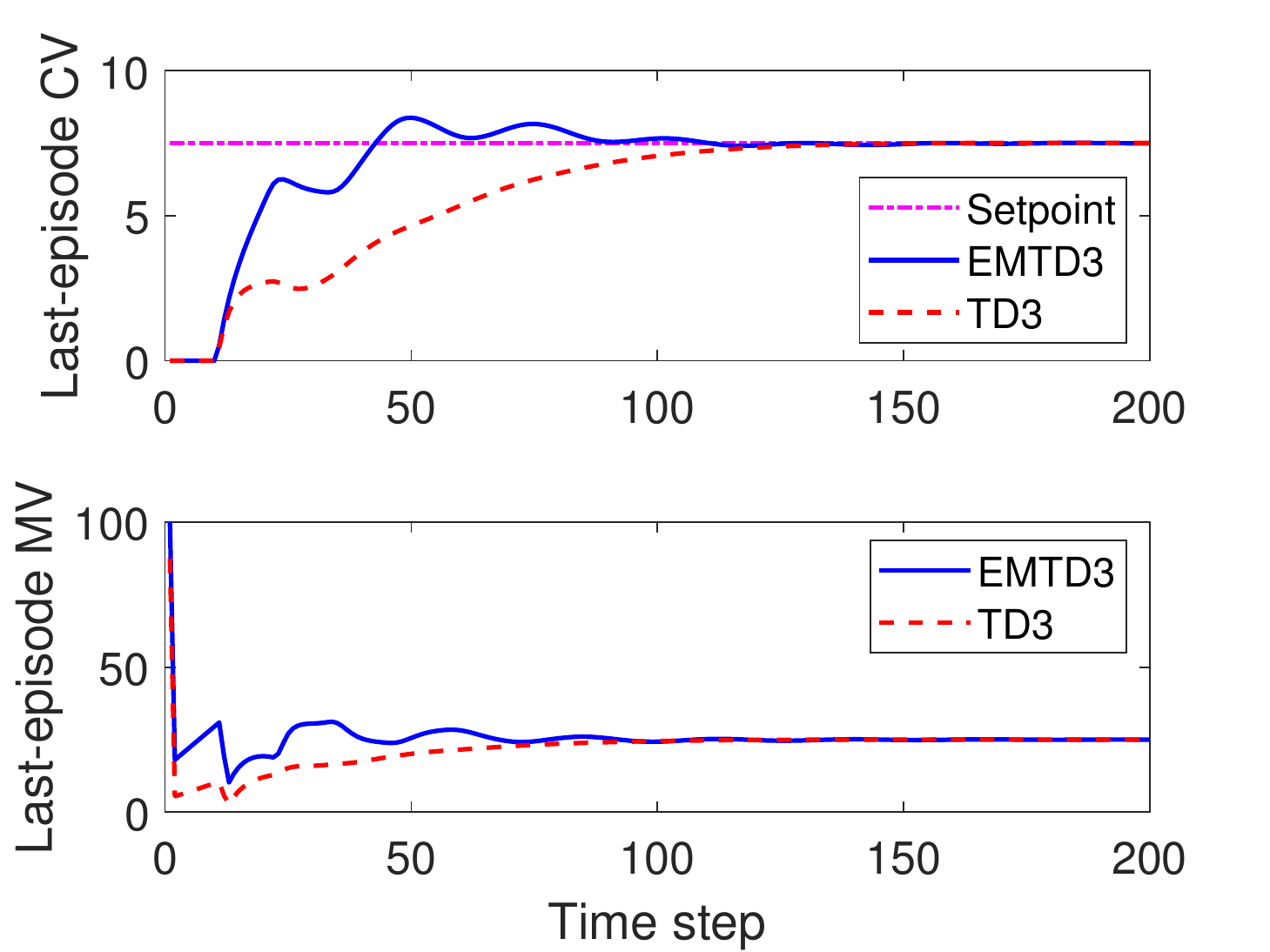}
	\caption{The PID control performance from TD3 and EMTD3 for Case II.}
	\label{fig: medium_action_cv}
\end{figure}

For this case study, we specify a much larger range for PID parameters than Case I to further assess the performance of EMTD3 and TD3 for PID tuning: $K_P\in [0,20]$, $\tau_{I}\in[0,20]$, and  $\tau_D\in[0,20]$. The exploration and exploitation schemes for both the TD3 and EMTD3 algorithms are the same as those in Case I. Fig. \ref{fig: medium_action exploration} illustrates the distribution of attempted PID parameters from these methods where the EMTD3 method again shows better exploration and thus converges closer to the optimum than TD3. Note that this action space is much larger and forms a more challenging problem than Case I. The learning curves in Fig. \ref{fig: medium_action} illustrate the convergence rate of the EMTD3 and TD3 algorithms. Similar to Case I, the EMTD3 method is more sample efficient and can deliver better ultimate reward than TD3, as shown in the inner plot in Fig. \ref{fig: medium_action}. The last-episode tracking performance based on the PID parameters delivered from EMTD3 and TD3 is shown in \ref{fig: medium_action_cv}, which indicates that the discovered PID parameters from EMTD3 can enable faster tracking performance with less tracking error. This is consistent with the observation with Fig. \ref{fig: medium_action exploration} where the solution from EMTD3 is much closer to the optimum than that from TD3. 


\section{Conclusion}
\label{sec: conclusion}
We proposed a sample-efficient entropy-maximizing TD3 method with an application to facilitate the automatic PID tuning. For the proposed method, an entropy-maximizing stochastic actor is adopted at the beginning to enable sufficient exploration, followed by a deterministic actor to focus on exploitation. We further developed a framework to formulate the PID tuning as a RL problem where the proposed EMTD3 is used to rapidly discover optimal PID parameters. Simulation case studies are provided to verify the sample efficiency and effectiveness of the proposed EMTD3 method in finding global optimum against the traditional TD3 method. Future work includes improving the algorithms to enable online adaptive PID tuning for nonlinear systems.

\addtolength{\textheight}{-12cm}   




\section*{ACKNOWLEDGMENT}
M. A. Chowdhury acknowledges the support of Distinguished Graduate Student Assistantships (DGSA) from the Texas Tech University. Q. Lu acknowledges the new faculty startup funds from the Texas Tech University.

\bibliographystyle{ieeetr}
\bibliography{scalab}

\end{document}